\documentclass[graybox]{svmult}

\usepackage{mathptmx}       
\usepackage{helvet}         
\usepackage{courier}        
\usepackage{type1cm}        
\usepackage{makeidx}         
\usepackage{graphicx}        
\usepackage{multicol}        
\usepackage[bottom]{footmisc}

\makeindex             

\begin{document}

\title*{Measurement Error Models in Astronomy}
\author{Brandon C. Kelly}
\institute{Brandon C. Kelly, Hubble Fellow \at Harvard-Smithsonian Center for Astrophysics, 60 Garden St, MS-06, \email{bckelly@cfa.harvard.edu}}
%
%
\maketitle

\abstract*{I discuss the effects of measurement error on regression and density estimation. I review the statistical methods that have been developed to correct for measurement error that are most popular in astronomical data analysis, discussing their advantages and disadvantages. I describe functional models for accounting for measurement error in regression, with emphasis on the methods of moments approach and the modified loss function approach. I then describe structural models for accounting for measurement error in regression and density estimation, with emphasis on maximum-likelihood and Bayesian methods. As an example of a Bayesian application, I analyze an astronomical data set subject to large measurement errors and a non-linear dependence between the response and covariate. I conclude with some directions for future research.}


\vspace{-1.5 in}
\abstract{I discuss the effects of measurement error on regression and density estimation. I review the statistical methods that have been developed to correct for measurement error that are most popular in astronomical data analysis, discussing their advantages and disadvantages. I describe functional models for accounting for measurement error in regression, with emphasis on the methods of moments approach and the modified loss function approach. I then describe structural models for accounting for measurement error in regression and density estimation, with emphasis on maximum-likelihood and Bayesian methods. As an example of a Bayesian application, I analyze an astronomical data set subject to large measurement errors and a non-linear dependence between the response and covariate. I conclude with some directions for future research.}


\section{Introduction}
\label{intro}

Measurement error is ubiquitous in astronomy. Astronomical data
consists of passive observations of objects, whereby astronomers are
able to directly measure the flux of an object as a function of
wavelength, its location on the sky, and the time of the
observation. Because the number of photons detected from an
astronomical object follows a Poisson process, this makes the
measurement of a source's intensity intrinsically subject to
measurement error, even if none is introduced from the
detector. Therefore, the very nature of astronomical data makes
measurement error unavoidable. Moreoever, quantities that are derived
from an object's observed emission, either by fitting a model to the
spectral energy distribution (SED) or by employing scaling
relationships, are also `measured' (derived) with error. Examples
include mass, metallicity, and distance. Often the measurement error
on the derived quantities is significant. This is unfortunate as
inference on the derived quantities is often the goal of astronomical
data analysis. Therefore, there has been considerable interest in how
to perform statistical inference in the presence of measurement
error. 

Measurement error is a problem that affects, at various levels, all
scientific research. Because of this, numerous methods for handling
measurement errors have been developed (Fuller 1987, Cheng \& Van Ness
1999, and Carroll et al. 2006 are good references). In this
contribution, I 
will present a survey of methods for handling measurement error that
have been developed and used in astronomical data analysis. Because
astronomical measurement errors are, in general, heteroskedastic
(having different variances), I will limit my discussion to methods
developed for heteroskedasticity. I will
focus on situations where a deterministic relationship is not assumed
between the variables, but where all variables of interest are random
and are measured with error. Because of this, I will ignore
situations where the measurement error
is the only source of randomness in one's data. An example of this type of situation
is fitting a model to an observed spectrum, where the measurement
error is the only source of randomness; i.e., in the absence of
measurement error a deterministic relationship is assumed between,
say, flux density and wavelength. Methods for handling measurement
error in this case are relatively well-established, and typically one
simply minimizes the usual $\chi^2$ statistic (e.g., Bevington
2003). However, it is worth pointing out that many complications may still
exist, and more sophisticated methods may be needed,
especially when dealing with low-count X-ray and $\gamma$-ray data
(e.g., van Dyk et al. 2001) or when incorporating calibration
unceratinties (Lee et al. 2011). Instead, I will focus on methods for
analyzing data from astronomical samples, where the variables are a
random sample from an underlying distribution. Within the context of
regression, this implies that intrinsic scatter (referred to as
equation error in the statistics literature) exists in the
relationship among the variables, and thus a deterministic
relationship is not assumed between the variables even without the
presence of measurement error.

Most of the techniques I will discuss focus on accounting for
measurement error in regression. The goal of regression is often to
understand how one variable changes with another. For example, how does the mass of a black hole change as a function of the stellar velocity dispersion of the host galaxy's bulge? Typically one
simply estimates how the average value and dispersion of one variable depends
on another. Measurement error statistical
models are typically divided into two types: `functional' and
`structural' models. In functional modeling, one assumes that the
unknown true values of the variables are fixed, whereas in structural
modeling the unknown true values of the variables have their own
intrinsic distribution. As a result, in structural modeling one must
parameterically model the distribution of the true values of the
variables, whereas in functional modeling one does not. Density
estimation is another important technique in astronomical data
analysis, being the foundation for luminosity and mass function
estimation. The methods I will discuss for handling measurement error
in structural models are also applicable to density estimation, as in this
case regression and density estimation are based on the same
formalism. When discussing regression methods, I will refer to the
`dependent' variable as the response, and the `independent variables'
as the covariates. 

\section{Notation and Error Model}

\label{s-emodel}

Throughout this work I will denote the measured response for the
$i^{\rm th}$ data point as $y_i$, and the measured covariate for the
$i^{\rm th}$ data point as $x_i$. I will denote the true values as
$\eta_i$ and $\xi_i$, respectively. If there are $p > 1$ covariates,
then I will use the vectors $\vec{x_i}$ and $\vec{\xi_i}$. I will use
$\vec{y}$ to denote the set of values of $y_i$ for each of the $n$
data points, $\vec{y} = [y_1, \ldots, y_n]$. To denote the set of
$x_i$, I will use $\vec{x} = {x_1, \ldots, x_n}$ if there is one
covariate, and the $n \times p$ matrix $X =
[\vec{x_1},\ldots,\vec{x_n}]$ if there are multiple covariates. I
assume the classical additive error models throughout this review,
unless otherwise specified:
\begin{eqnarray}
  \eta_i & = & f(\vec{\xi_i}, \theta) + \epsilon_i \\
  \vec{x_i} & = & \vec{\xi_i} + \vec{\epsilon_{x,i}} \\
  y_i & = & \eta_i + \epsilon_{y,i}.
\end{eqnarray}
The function $f(\vec{\xi},\theta)$ describes how the mean value of
$\eta$ depends on $\vec{\xi}$ as a function of the parameters,
$\theta$. For example, for linear regression $f(\vec{\xi}, \theta) =
\alpha + \vec{\beta}^T \vec{\xi}$ with $\theta = (\alpha, \beta)$
denoting the slopes and intercept. The
terms $\epsilon_i, \vec{\epsilon_{x,i}},$ and 
$\epsilon_{y,i}$ are random variables denoting the intrinsic scatter
in $\eta$ at fixed $\xi$ (i.e., the equation error), the measurement
error in $\vec{x_i}$, and the measurement error in $y_i$,
respectively. The random variables $\epsilon_i, \vec{\epsilon_{x,i}},$
and $\epsilon_{y,i}$ are assumed to have zero mean and variances
$Var(\epsilon_i) = \sigma^2, Var(\epsilon_{y,i}) = \sigma^2_{y,i},$
and $Var(\vec{\epsilon_{x,i}}) = \Sigma_{\vec{x},i}$. As is typical in
astronomy, the parameter $\sigma^2$ is assumed to be unknown and a
free parameter in the model, while the variances in the measurement
errors, $\sigma^2_{y,i}$ and $\Sigma_{\vec{x},i}$, are assumed known. The measurement errors
are assumed to be independent of $\epsilon_i$. In addition, for
simplicity I also assume that the measurement errors in $y_i$ and
$\vec{x_i}$ are independent, unless otherwise specified. However, this is not always true, and
many methods are able to handle correlated measurement errors, see the
references for individual techniques for further details.

Following Gelman et al. (2004), I will also typically use the notation
$p(\cdot)$ to denote the probability density of the argument. For
example, $p(x)$ denotes the marginal probability density of $x$,
$p(y|x)$ denotes the conditional probability density of $y$ given $x$,
and $p(y,x)$ denotes the joint probability density of $y$ and $x$. It
should be understood that $p(\cdot)$ will not always have the
same functional form, and that this must be inferred from context,
i.e., it is not necessarily true that $p(x) = p(y)$ even if $x =
y$. When this may be confusing, I use different symbols
to denote different probability densities.

\section{Effects of Measurement Error}

\label{s-effects}


Measurement error has the effect of blurring and broadening the
distribution of quantities, similar to the blurring of astronomical
images by a point spread function. This makes statistical inference based
on the measured values biased, and smears out any trends in the
data. The distribution of the measured quantities is obtained as
\begin{equation}
  p(y,\vec{x}) = \int \int p(y,\vec{x}|\eta,\vec{\xi}) p(\eta,\vec{\xi})\ d\eta\ d\vec{\xi}.
  \label{eq-pyx}
\end{equation}
Under the additive error model of \S~\ref{s-emodel}, Equation
(\ref{eq-pyx}) simplifies to
\begin{equation}
  p(y,\vec{x}) = \int f(y - \eta) \int g(\vec{x} - \vec{\xi}) p(\eta,\vec{\xi}) \ d\vec{\xi}\ d\eta,
  \label{eq-pyx_conv}
\end{equation}
where $f(\cdot)$ and $g(\cdot)$ denote the probability distributions of
the measurement errors $\epsilon_{y}$ and $\epsilon_x$,
respectively. Equation (\ref{eq-pyx_conv}) shows that under additive
measurement error, the observed distribution of a set of quantities is
the convolution of the intrinsic distribution with the measurement
error distribution. Convolution has the effect of broadening
distributions, which biases density estimation and masks trends.

\begin{figure}[t]
\includegraphics[scale=0.21,angle=0]{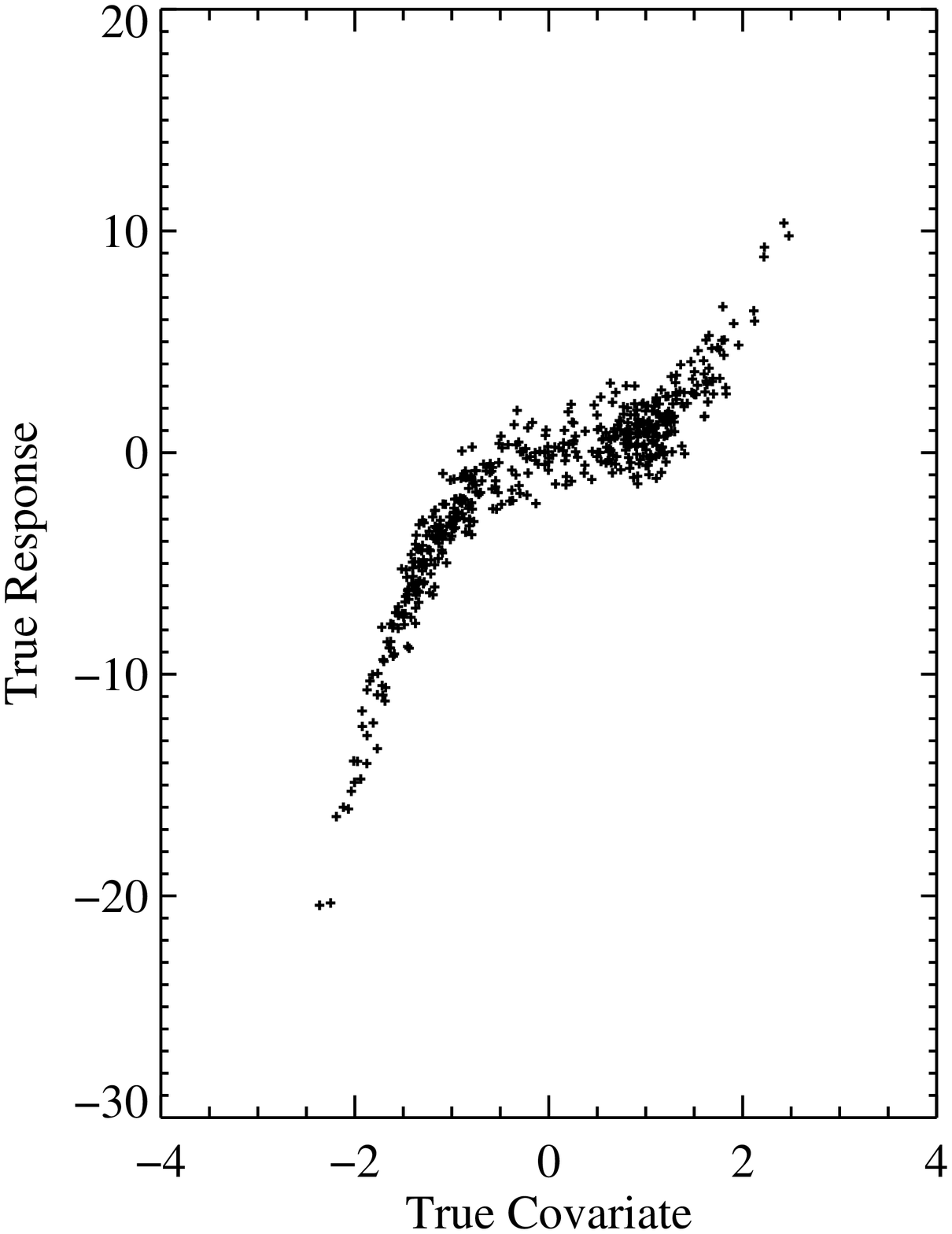}
\includegraphics[scale=0.21,angle=0]{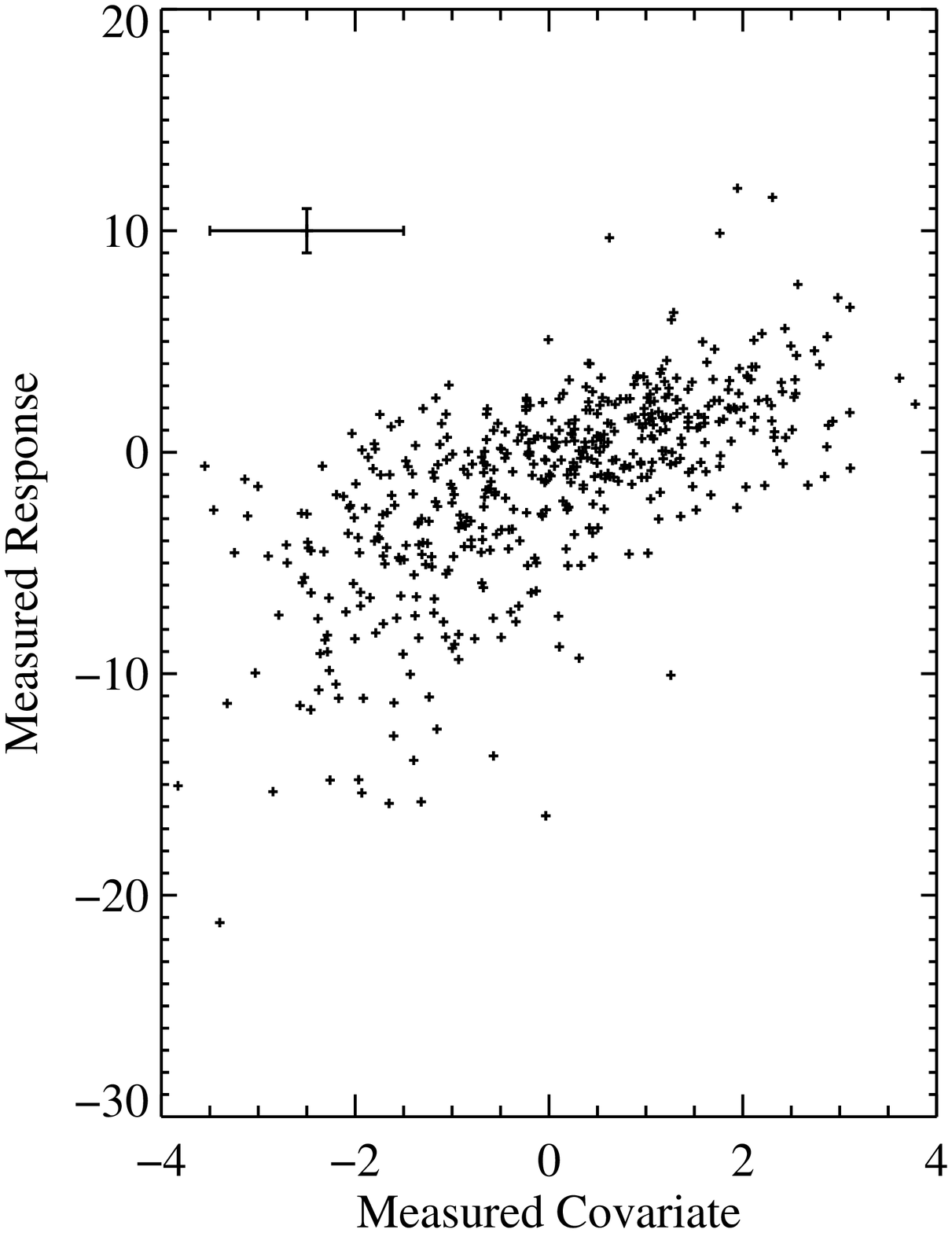}
\includegraphics[scale=0.21,angle=0]{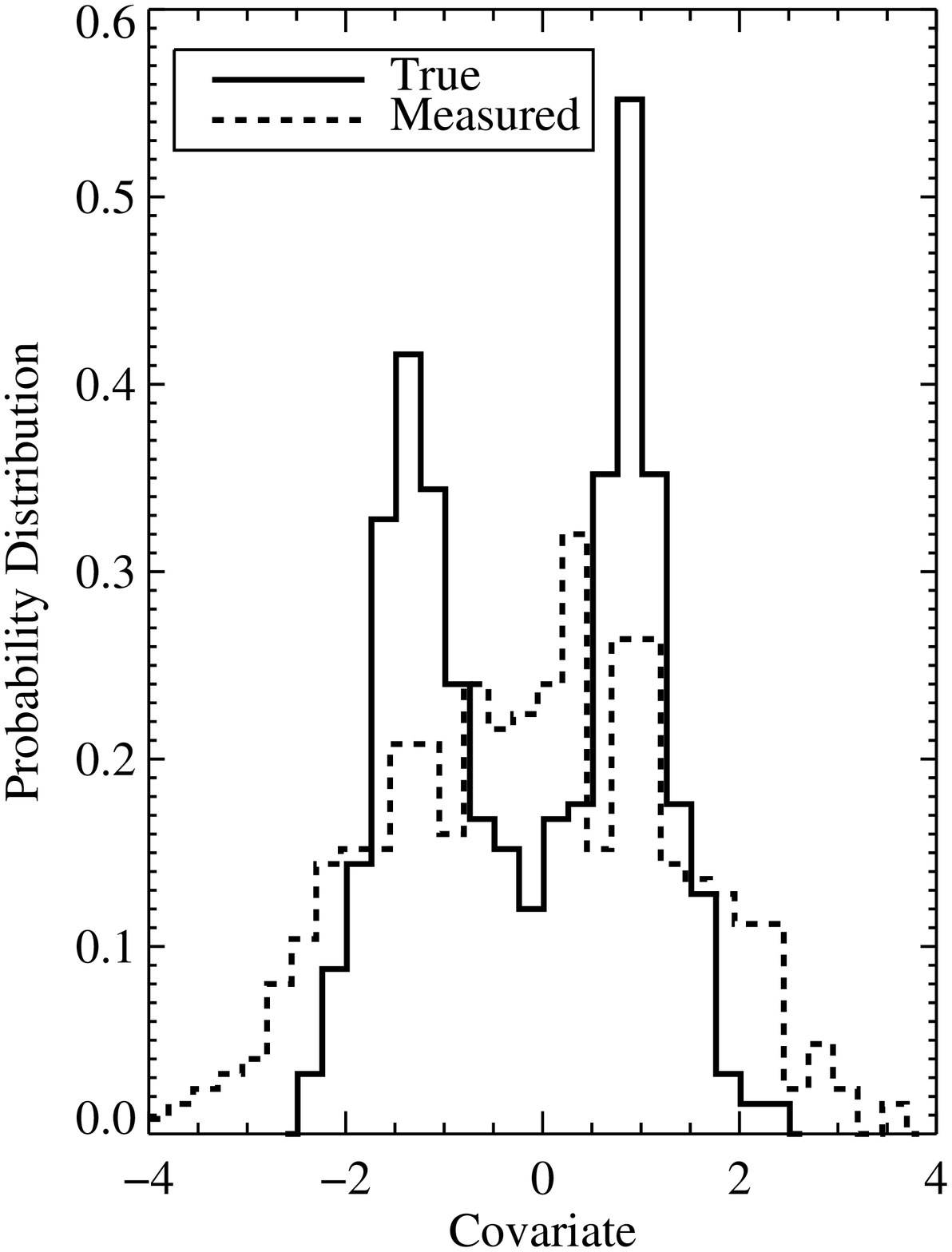}
%
%
\caption{Illustration of the effect of measurement error on regression
  and density estimation, using a simulated sample. The true distribution of the response and covariate (left),
  compared with the measured distribution (center). The error bars in
  the center plot denote the standard deviation of the Gaussian
  measurement errors. The measurement errors have effectively washed
  out any visual evidence for a tight non-linear relationship between
  the response and covariate. The right plot shows the distribution of
the true and measured values of the covariate. The measurement errors
have washed out any evidence for bimodality in the distribution, and
significantly broadened it.}
\label{f-measerr_effects}       
\end{figure}

Some of the effects of measurement error are illustrated in Figure
\ref{f-measerr_effects}. Here, I simulated a sample of covariates from
a bimodal distribution, and simulated the response assuming a
nonlinear relationship between $\eta$ and $\xi$. I then added large
measurement error to both $\eta$ and $\xi$. As can be seen,
measurement error has blurred out many of the features in the data
set, and broadened the distributions.

%

To further see how measurement
error biases statistical inference for regression, consider the additive error model for
linear regression, assuming one covariate. In addition, for simplicity
assume that
the measurement errors are homoskedastic (having the same variance)
for both the response and covariate. If one were
to ignore measurement error and proceed through the usual ordinary
least-squares (OLS) analysis, then one would obtain the following estimates
for the slope, variance in the intrinsic scatter, and uncertainty in
the estimated slope (assume the intercept, $\alpha$, is known):
\begin{eqnarray}
  \hat{\beta}_{OLS} & = &\frac{Cov(x,y)}{Var(x)} =
  \frac{Cov(\xi,\eta)}{Var(\xi) + \sigma^2_x} 
  \label{eq-beta_ols} \\
  \hat{\sigma}^2_{OLS} & = & Var(y - \alpha - \hat{\beta}_{OLS} x) =
  (\beta^2 - \hat{\beta}^2_{OLS}) Var(\xi) + \hat{\beta}^2_{OLS}
  \sigma^2_x + \sigma^2_y + \sigma^2 \label{eq-sigsqr_ols} \\
  Var(\hat{\beta}_{ols}) & = & \frac{\hat{\sigma}^2_{OLS}}{Var(x)} =
  \frac{\hat{\sigma}^2_{OLS}}{Var(\xi) + \sigma^2_x},
  \label{eq-slope_err_ols}
\end{eqnarray}
where $\beta$ and $\sigma^2$ are the true values of the slope and
variance in intrinsic scatter. From Equations
(\ref{eq-beta_ols})--(\ref{eq-slope_err_ols}) we can deduce the following:
\begin{itemize}
\item
  Equation (\ref{eq-beta_ols}) shows that measurement error in the covariate attenuates the regression slope, biasing it toward
  zero. Therefore, trends between the response and the covariate will
  appear weaker than they really are. If the measurement error in the
  covariate is negligible, then there is no bias in the slope even if
  the measurement errors in the response are large.
\item
  Equation (\ref{eq-sigsqr_ols}) shows that measurement error in both the response and covariate bias the
  estimate of $\sigma^2$ upward. Therefore, the variance in the
  response about the regression line will appear larger than it really
  is.
\item
  Equation (\ref{eq-slope_err_ols}) show that measurement error in the covariate causes one to underestimate the
  error in the estimated slope. Thus, if the covariate is
  significantly contaminated by measurement error, then one would
  incorrectly conclude that the slope is precisely estimated to be
  $\approx 0$, and therefore conclude that there is no relationship between the
  response and covariate!
\end{itemize}
Clearly measurement error can have a significant effect on one's data
analysis, and ignoring it can lead to erroneous conclusions. Luckily,
a number of statistical methods have been developed for handling measurement errors.

\section{Functional Methods for Accounting for Measurement Error in Regression}
\label{s-funcmod}

A variety of functional models have been proposed for handling measurement
errors in regression, and here I summarize the methods that are commonly used in the
astronomical literature. Since heteroskedastic measurement errors are the
norm in astronomy, I only discuss methods that allow the variances in
the measurement error to vary among the observations. Moreover, as
discussed earlier, I focus on methods that incorporate intrinsic
scatter in the relationship between the response and
covariate. The
reader is referred to Carroll et al. (2006) for a more thorough and
general discussion of methods developed for handling measurement
error. 

\subsection{Method of Moments Approach for Linear Regression}
\label{s-mm}

In linear regression the least-squares estimates of the intercept,
slope, and intrinsic dispersion are obtained from the moments of the
data. In the previous section I showed that the moments of the
observed data are biased estimates of the moments of the intrinsic
distribution when the data are measured with error. Therefore a simple
method of accounting for measurement error in linear regression is to
estimate the moments of the true values of the data, and then use
these estimated moments to estimate the regression parameters. This is
the idea behind the method of moments (MM) estimators, where the moments of the
observed data are `debiased' by removing the contribution from the
measurement errors.

Akritas \& Bershady (1996) describe a methods of moments approach for linear regression that
handles heteroskedastic measurement error in both the response and
covariate, intrinsic scatter, and correlation between the response and
covariate measurement error. Akritas \& Bershady used their method to characterize the color-luminosity and Tully-Fisher relationships for galaxies. Their estimators, as is typical for the method
of moments, assume the additive error model of \S~\ref{s-emodel} with
the mean value of $\eta$ depending linearly on $\xi$: $f(\xi,\theta) = \alpha + \beta \xi$.
They do not assume a particular distribution for the measurement
errors, the covariate, or the intrinsic scatter. However, their approach does assume that the
variance in the measurement errors and correlation between the
measurement errors are known. They call their estimator the BCES
estimator, for bivariate correlated errors and intrinsic
scatter. 

Denote the covariance between the measurement errors in the
response and covariate as $Cov(\epsilon_{y,i},\epsilon_{x,i}) =
\sigma_{yx,i}$. Also, denote the sample average for $x$ as
$\bar{X}$, the sample average for $y$ as $\bar{Y}$, the sample
variance for $x$ as $V_x$, the sample variance for $y$ as $V_y$, and
the sample covariance between $x$
and $y$ as $V_{xy}$. Then, the methods of moments estimators are
\begin{eqnarray}
\hat{\beta}_{MM} & = & \frac{V_{xy} - \bar{\sigma}_{yx}}{V_x -
  \bar{\sigma}^2_{x}} \label{eq-mm_slope} \\
\hat{\alpha}_{MM} & = & \bar{Y} - \hat{\beta}_{MM}
\bar{X} \label{eq-mm_intercept}
\end{eqnarray}
where $\bar{\sigma}_{yx} = \sum_{i=1}^n \sigma_{yx,i} / n$ and
$\bar{\sigma}^2_x = \sum_{i=1}^n \sigma^2_{x,i} / n$. Akritas \&
Bershady (1996) show that the MM estimators are asymptotically
unbiased, that the sampling distribution of the MM estimators is
asymptotically normal, and describe how to estimate the asymptotic
covariance matrix of $\hat{\alpha}_{MM}$ and
$\hat{\beta}_{MM}$. Patriota et al. (2009) derive the asymptotic
covariance matrix of the MM estimators under the additional assumption that
the measurement errors are normally distributed, creating more
powerful hypothesis testing when this is true. In addition, Cheng \&
Riu (2006) give the MM estimator for the variance in the intrinsic
scatter: 
\begin{equation}
\hat{\sigma}^2_{MM} = V_y - \hat{\beta}_{MM} (V_{xy} - \bar{\sigma}_{yx}) - \bar{\sigma}^2_y,
\label{eq-mm_intvar}
\end{equation}
where $\bar{\sigma}^2_y$ is the sample average of $\sigma^2_{y,i}$.

The main advantage of the MM estimators are that they do not make any
assumptions about the distribution of the measurements errors, about
the distribution of the covariate, nor about the distribution of the intrinsic scatter. This is attractive, is it makes
the MM estimators robust. One of the disadvantages of the MM
estimators is that they are not as precise as some other methods, such as structural models, when the distributions of
$\epsilon_{x}, \epsilon_{y}, \epsilon,$ and $\xi$ are known, or at
least when they can be accurately modeled parameterically, as the MM estimators do not impose prior assumptions about the distributions. Another
disadvantage is that the MM estimators tend to be highly variable when
the sample size is small, and/or the measurement errors are
large. This is on account of the term $V_x - \bar{\sigma}^2_x$ in the
denominator of the equation for $\hat{\beta}_{MM}$. When the sample
size is small, then $V_x$ is more variable, and it is possible that
$V_x \sim \bar{\sigma}^2_x$.  This is also possible when measurement
errors are large, as the variance in $x$ becomes dominated by the
measurement errors. When this occurs, the estimate for the slope can
become very large, or change sign. Similarly, if the measurement
errors in $y$ are large, then the MM estimator for the intrinsic
dispersion can become negative, which is impossible. Therefore,
despite the robustness of the MM estimators, more stable estimators
should be used when the sample size is small, or when the measurement
errors make up a significant component to the variance in the data. 

\subsection{Modified Loss Function Approach}
\label{s-corrscore}

Modified loss function methods modify the figure of merit function (i.e., the `loss' function), to incorporate 
measurement error. The weighted squared error loss function is the most common loss
function used in astronomy. A weighted least squares (WLS) estimator
for linear regression was proposed by Sprent (1966) to minimize the
following loss function for the special case of no intrinsic scatter: 
\begin{equation}
  Q_{WLS}(\alpha,\beta) = \sum_{i=1}^n \frac{(y_i - \alpha - \beta
    x_i)^2}{\sigma_{y,i}^2 + \beta^2 \sigma_{x,i}^2}. 
  \label{eq-wls}
\end{equation}
The weights in Equation (\ref{eq-wls}) reflect the contribution of the
measurement errors to the squared
error. Here I have used the notation $Q_{WLS}(\alpha,\beta)$ instead
of the more commonly used $\chi^2$ to emphasize the fact that Equation
(\ref{eq-wls}) is a loss (or figure of merit) function, and will not
necessarily follow a $\chi^2$ distribution even if the errors are Gaussian (although one can still use Equation (\ref{eq-wls}) regardless of the distribution of the measurement errors). Note that this implies that
one cannot derive uncertainties in the parameters by looking for
regions of constant $\Delta Q_{WLS}(\alpha,\beta)$. As with the method
of moments estimators, the WLS estimators do not make any assumptions
about the distribution of the measurement errors, covariate, or
intrinsic scatter. 

The loss function defined by Equation (\ref{eq-wls}) assumes that
there is no intrinsic scatter in the relationship between the response
and covariate. How then to modify Equation (\ref{eq-wls}) to include
the intrinsic scatter? Motivated by their work on characterizing the $M_{BH}$--$\sigma_*$ relationship, Tremaine et al. (2002) suggested using the
following modified WLS loss function: 
\begin{equation}
  \tilde{Q}_{WLS}(\alpha, \beta,\sigma^2) = \sum_{i=1}^n \frac{(y_i -
    \alpha - \beta x_i)^2}{\sigma^2 + \sigma_{y,i}^2 + \beta^2
    \sigma_{x,i}^2}. 
    \label{eq-fitexy}
\end{equation}
While the addition of $\sigma^2$ to the denominator of
Equation (\ref{eq-fitexy}) is intuitive, as it reweights the loss
function to incorporate the intrinsic scatter, the unknown value of
$\sigma^2$ creates difficulties for the WLS estimator based on
$\tilde{Q}_{WLS}(\alpha, \beta,\sigma^2)$. As discussed in Kelly (2007),
Equation (\ref{eq-fitexy}) can only be minimized with respect to
$\alpha$ and $\beta$ at fixed $\sigma^2$, as the minimum of Equation
(\ref{eq-fitexy}) occurs at $\sigma^2 \rightarrow \infty$ for any
value of $\alpha$ and $\beta$. Clearly, one cannot estimate the
regression parameters by minimizing
$\tilde{Q}_{WLS}(\alpha,\beta,\sigma^2)$. Instead, the most common approach (as
suggested by Tremaine et al. (2002)) is to initially use $\sigma^2 =
0$, and then find the values of $\alpha$ and $\beta$ which
minimize Equation (\ref{eq-wls}). Then, using these best-fit values
for $\alpha$ and $\beta$, $\sigma^2$ is estimated by finding the value
such that $\tilde{Q}_{WLS}(\alpha,\beta,\sigma^2) / (n - 2) =
1$. Unfortunately, the properties of the WLS estimator based on this
procedure, such as its bias and asymptotic distribution, are
unknown. Kelly (2007) performed simulations to study the behavior of
the WLS estimator based on Equation (\ref{eq-fitexy}) when the data
are contaminated by large measurement error, and compared with the MM
estimator and a maximum-likelihood estimator (see \S~\ref{s-likmeth}). In general, the WLS
estimator gave biased values for the slope, while the MM estimator for the
slope was approximately unbiased except in the limit of extreme
measurement error, and the maximum-likelihood estimator was
approximately unbiased except in the limit of a small sample with
extreme measurement error. Therefore, based on the problems associated
with the WLS estimator based on Equation (\ref{eq-fitexy}), I do not
recommend its use. 

While the modification to the least squares loss function by Equation
(\ref{eq-fitexy}) exhibits some problems, it is still possible to derive
consistent estimators for the regression parameters by modifying the
least squares loss function. Instead, consider the following modified
loss function: 
\begin{equation}
  Q(\alpha,\beta,\sigma^2) = \frac{1}{\sigma^2} \sum_{i=1}^n \left[
    (y_i - \alpha - \beta x_i)^2 - \sigma^2_{y,i} - \beta^2
    \sigma^2_{x,i} \right]. 
  \label{eq-mmloss}
\end{equation}
Equation (\ref{eq-mmloss}) corrects the usual least-squares loss
function by subtracting off the contribution to the squared error from the
measurement errors, and is therefore an estimate of the loss function
that would have been obtained if there was no measurement error. Minimization of Equation (\ref{eq-mmloss}) with
respect to $(\alpha,\beta,\sigma^2)$ results in the MM estimators
given by Equations (\ref{eq-mm_slope})--(\ref{eq-mm_intvar}) (Cheng \&
Van Ness 1999). Therefore, the method of moments estimators can be
understood as resulting from a corrected least squares loss function. 

Thus far I have focused on linear regression. However, there are cases
where a non-linear relationship may exist between the average value of
the response and the covariate, and one desires to use a functional
model. Patriota \& Bolfarine (2008) describe 
a corrected score method for polynomial regression under the
heteroskedastic additive
error model (\S~\ref{s-emodel}), which they applied to an astronomical
data set. The reader is referred to their work for further details.

\section{Structural Methods for Regression and Density Estimation}

Structural models for regression are those that make assumptions about the
distribution of the covariate. As such, they are also applicable to
density estimation. I will focus on structural models that rely on the
construction of a likelihood function\footnote{The
  likelihood function is the probability of observing the data, given
  some parameters. It requires assuming a parameteric form for the
  sampling distribution of the data.}, therefore requiring one to
specify a parameteric model for the distributions of the measurement
errors, intrinsic scatter, and covariates. These methods include both
maximum-likelihood estimators and Bayesian methods. Likelihood-based
techniques have the advantage that they are flexible and may be
applied to a variety of problems, including those requiring non-linear
forms for $f(\xi,\theta)$, variance in intrinsic scatter that depends
on the covariate, and data sets that include censoring\footnote{Data
  are said to be censored when only an upper or lower limit is
  available.} and truncation. However, they have the disadvantages
that they are computationaly expensive, and that one must assume a
parameteric form for all distributions involved, decreasing their
robustness. That being said, it is possible to use highly flexible
parameteric forms, increasing the robustness of likelihood based
methods (Huang et al. 2006). Moreover, the additional assumptions
involved in the parameteric modeling typically buys one an increase in
efficiency, providing smaller standard errors for the
maximum-likelihood and Bayesian estimators when the parameteric
statistical model is a good description of the data. 

\subsection{Constructing the Likelihood Function}
\label{s-likmeth}

The basic idea behind likelihood-based methods is to treat the
measurement errors as a missing data problem. Little \& Rubin (2002)
describe methods for handling missing data, while Gelman et
al. (2004) describe Bayesian approaches to the missing data
problem. First, one formulates the likelihood function for
the complete data, i.e., the likelihood function for both the measured
and true values of the data. In general, for regression we have the
following hierarchical model: 
\begin{eqnarray}
  \vec{\xi}_i & \sim & p(\vec{\xi}|\psi) \label{eq-xidist} \\
  \eta_i | \vec{\xi}_i & \sim & p(\eta|\vec{\xi},\theta) \label{eq-etadist} \\
  y_i,\vec{x}_i | \eta_i, \xi_i & \sim & p(y,\vec{x}|\eta,\vec{\xi}). \label{eq-measdist}
\end{eqnarray}
The notation $z \sim p(z)$ means that the random variable $z$ is drawn
from the probability distribution $p(z)$. The distributions
$p(\vec{\xi}|\psi), p(\eta|\vec{\xi},\theta),$ and
$p(y,\vec{x}|\eta,\xi)$ are the distributions for the covariates, the
response given the covariate, and the measured data, respectively. The
distribution for the covariate is parameterized by $\psi$, while the
distribution for $\eta$ at a given $\vec{\xi}$ is parameterized by
$\theta$; note that here I have absorbed the parameter describing the
variance in the intrinsic scatter into $\theta$, whereas in the
previous sections I have kept $\sigma^2$ seperate from $\theta$. For
simplicity, I assume that the distribution of the measurement errors
is considered known, as is typically the case in astronomy. If
additional parameters are needed to describe the distribution of the
measured data, e.g., if the variance in the measurement errors is
unknown, then these should be included in Equation
(\ref{eq-measdist}). Most of the interest in regression lies in
inference on $\theta$, which describes how the response depends on the
covariates. If, instead of regression we are interested in density
estimation, then there is no response variable and only Equations
(\ref{eq-xidist}) and (\ref{eq-measdist}) are used.

Under the statistical model given by Equations
(\ref{eq-xidist})--(\ref{eq-measdist}), the complete data likelihood
function for the $i^{\rm th}$ data point is 
\begin{equation}
  p(y_i,\vec{x}_i,\eta_i,\vec{\xi}_i|\theta,\psi) =
  p(y_i,\vec{x}_i|\eta_i,\xi_i) p(\eta_i|\vec{\xi}_i,\theta)
  p(\vec{\xi}_i|\psi). 
  \label{eq-complik}
\end{equation}
In order to calculate the observed data likelihood function for the
$i^{\rm th}$ data point, we integrate out the missing (and thus
unknown) data from the complete data likelihood function: 
\begin{equation}
  p(y_i,\vec{x}_i|\theta,\psi) = \int \int
  p(y_i,\vec{x}_i|\eta,\vec{\xi}) p(\eta|\vec{\xi},\theta)
  p(\vec{\xi}|\psi)\ d\eta\ d\vec{\xi} \label{eq-obslik} 
\end{equation}
When the data points are statistically independent, as is almost
always the case, the observed data likelihood function for the entire
data set is the product of Equation (\ref{eq-obslik}) over the $i = 1,
\ldots, n$ data points. Further details on this procedure can be found
in Carroll et al. (2006). Once one has chosen parameteric forms for
the distributions involved in Equations
(\ref{eq-xidist})--(\ref{eq-measdist}), one can use Equation
(\ref{eq-obslik}) to compute the maximum-likelihood estimate for the
parameters $(\theta,\psi)$ and use the likelihood ratio to estimate
confidence regions for the parameters. That's it! Of course, in
practice this is not so simple, as computing the integrations involved
in Equation (\ref{eq-obslik}) and performing the optimization of Equation
(\ref{eq-obslik}) can be numerically difficult. The
Expectation-Maximization (EM) algorithm is often helpful,
and additional numerical techniques are described in, for example,
Press et al. (2007) and Robert \& Casella (2004). 

As an example of the likelihood approach, consider the following
simple model. Assume the measurement errors to be normally distribution with zero mean and known variances, as described in
\S~\ref{s-emodel}. For the regression model, assume that the response ($\eta$) at fixed covariate ($\vec{\xi}$) is normally
distributed with mean $f(\vec{\xi},\theta) = \alpha + \beta^T \vec{\xi}$
and variance $\sigma^2$; this is the usual linear regression model
with Gaussian intrinsic scatter. The distribution of the covariates is
assumed to be a $p$-dimensional multivariate normal density with mean
$\mu$ and covariance matrix $T$. Under this model, the parameters are
$\theta = (\alpha, \beta,\sigma^2)$ and $\psi =
(\mu,T)$. For this model, the integrals in Equation (\ref{eq-obslik})
can be done analytically. Denoting $\vec{z}_i = (y_i, \vec{x}_i)$, the
measured data likelihood is 
\begin{eqnarray}
  p(\vec{y},X|\theta,\psi) & = & \prod_{i=1}^{n} \frac{1}{(2\pi)^{(p+1)/2} 
    |V_{i}|^{1/2}} \exp \left\{ -\frac{1}{2} (\vec{z}_i - \zeta)^T V_{i}^{-1} 
    (\vec{z}_i - \zeta) \right\} \label{eq-mobslik} \\
  \zeta & = & (\alpha + \beta^T \mu, \mu) \label{eq-mzeta} \\
  V_{i} & = & \left( \begin{array}{cc}
      \beta^T T \beta + \sigma^2 + \sigma^2_{y,i} & 
      \beta^T T \\
      T \beta & T + \Sigma_{x,i} \end{array} \right).
  \label{eq-msigmaz}
\end{eqnarray}

The Gaussian likelihood model described here is commonly used, but it
is not robust and can be subject to considerable systematic error due
to model mispecification (e.g., Huang et al. 2006). Motivated by this,
several authors have proposed using a mixture of Gaussian functions as
a model for the distribution of the covariates (e.g., Carroll et
al. 1999, Roy \& Banerjee 2006, Kelly 2007). Bovy et al. (2009)
describe a mixture of Gaussian functions model for density estimation
when some of the measurements are missing at random. Kelly et
al. (2008) describe a mixture of Gaussian functions model for density
estimation of a truncated sample, with emphasis on luminosity
function estimation. The mixture of
Gaussian functions model inherits much of the mathematical simplicity
of the Gaussian model, enabling an analytic calculation of the
observed data likelihood, while still being flexible enough to model
most realistic astrophysical distributions. In addition, Andreon (2006)
describe a model for incorporating contamination from a background
distribution, and model the distribution of the covariates as a
mixture of Schechter functions\footnote{The Schechter function is an
  unnormalized Gamma distribution. It is commonly used in astronomy as
  a model for the number density of galaxies in the universe as a
  function of their luminosity.}.  

\subsection{Bayesian Methods and an Example}
\label{s-bayes}

Bayesian methods build on the likelihood methods described in
\S~\ref{s-likmeth} and compute the probability distribution of the
parameters, given the observed data; this is called the `posterior'
distribution. This is done by first assuming a `prior' distribution on
the parameters, $p(\theta,\psi)$, where the prior distribution
quantifies our information on the parameters $\theta$ and
$\psi$ before we take any of the data. The posterior distribution is then related to the prior and
the likelihood by 
\begin{equation}
  p(\theta,\psi|\vec{y},X) = p(\theta,\psi) p(\vec{y},X|\theta,\psi) \label{eq-posterior}.
\end{equation}
For example, for the Gaussian model described by Equations
(\ref{eq-mobslik})--(\ref{eq-msigmaz}), and assuming a
uniform prior on the parameters ($p(\alpha,\vec{\beta},\sigma^2,\mu,T) \propto
1$\footnote{Technically this is uniform subject to the conditions that
  $\sigma^2 > 0$ and $|T| > 0$.}), the posterior distribution for
$(\alpha, \vec{\beta}, \sigma^2, \mu,T)$ is proportional to Equation
(\ref{eq-mobslik}) as a function of these parameters. Bayesian methods differ from the
frequentist likelihood methods, such as maximum-likelihood, in that
the inclusion of the prior distribution enables one to calculate the
probability of the parameters, given the observed data. This implies
that, in theory, the posterior distribution is exact, and therefore
uncertainties on the parameters are reliable and easy to interpret
regardless of the sample size and complexity of the statistical
model. In contrast, the maximum-likelihood methods compute a point
estimate of the parameters, and then use various methods (e.g., the
likelihood ratio or bootstrap) to estimate the sampling distribution
of the parameters, from which confidence regions are derived. The
maximum-likelihood methods are useful, but it can become difficult to
estimate the sampling distribution when the sample size is small, or
for highly complex and difficult models. 

Bayesian methods have become increasingly popular in astronomy, as
well as in other scientific disciplines. The primary driver of this
increase in popularity has been the advancements in statistical
computing that have enabled Bayesian inference, namely the use of
Markov Chain Monte Carlo (MCMC) methods. Details of MCMC methods may be found in
Gelman et al. (2004) and Liu (2004), and for an example of an MCMC algorithm
under linear regression and heteroskedastic measurement errors, see
Kelly (2007). One of the primary advantages of MCMC methods is that
they are modular, and we can divide the computational problem
up into smaller computational problems that are easier to
solve. Because the true values of the data are not known,
they are treated as additional parameters, and thus can also be
updated via MCMC. We can also incorporate upper and lower limits in a
straightforward manner through this approach by treating their true values as missing data (Kelly 2007), although the definition of upper limit in astronomy is not always straightforward (Kashyap et al. 2010). These
properties of MCMC samplers are a significant advantage of the Bayesian
approach, as we avoid the integration over the true values of the data required in Equation
(\ref{eq-obslik}) for the maximum-likelihood approach, and we obtain
improved estimates for the true values of the data. In fact, often it is
easier to program a MCMC sampler and perform Bayesian inference than
it is to do the optimization and numerical integration required for
maximum-likelihood.

As an illustration of the Bayesian approach, I consider a data set
from Constantin et al. (2011, in prep) comparing the X-ray photon
index, $\Gamma_X$, with the luminosity relative to the Eddington limit
(i.e., the Eddington ratio, $L / L_{Edd}$) for a sample of Active
Galactic Nuclei (AGN)\footnote{AGN are believed to be supermassive
  black holes that are accreting gas and are located in the center of
  a galaxy.}. The measured data are shown in Figure 2a. The X-ray
photon index provides a measure of how much energy is being 
released through soft X-rays as opposed to hard X-rays, and the
Eddington Luminosity is the luminosity at which outward radiation and
inward gravitational pressure balance for a spherical geometry. This data set
provides a good illustration of the power of the Bayesian approach, as
the average value of the response exhibits a non-linear and
non-monotonic dependence on the covariate, and the measurement errors
are very large in both the response and covariate. The values of the
Eddington ratio (i.e., the covariate) where 
the X-ray photon index (i.e., the response) changes its dependence on
$L / L_{Edd}$ are of
particular interest, as models of black hole accretion flows suggest
that the accretion flow geometry changes at certain critical values of
the Eddington ratio. Because of this, and the non-linear appearance in
the data, I have chosen to model the data using a segmented line
with two knots, where the slope of the line changes at
the knots. I modeled the intrinsic distribution of $\log L_X /
L_{Edd}$ as a mixture of three Gaussian distributions. To make the
model robust against outliers, I assume that 
the both measurement errors and the intrinsic scatter follow a
Student's t-distribution with eight and four degrees of freedom,
respectively. I used the MCMC algorithms described in
Chapter 9 of Carroll et al. (2006) and Kelly (2007) as the basis for
my MCMC sampler under this model, and include an ancillarity-sufficiency
interweaving strategy for increased efficiency (Yu \& Meng 2011). This MCMC
algorithm produces both random draws of the parameters for the
segmented line model from their posterior distribution, but also random draws of
the true values of the Eddington ratio and photon index from their
posterior distribution. 

%
\begin{figure}[t]
\includegraphics[scale=0.25, angle=90]{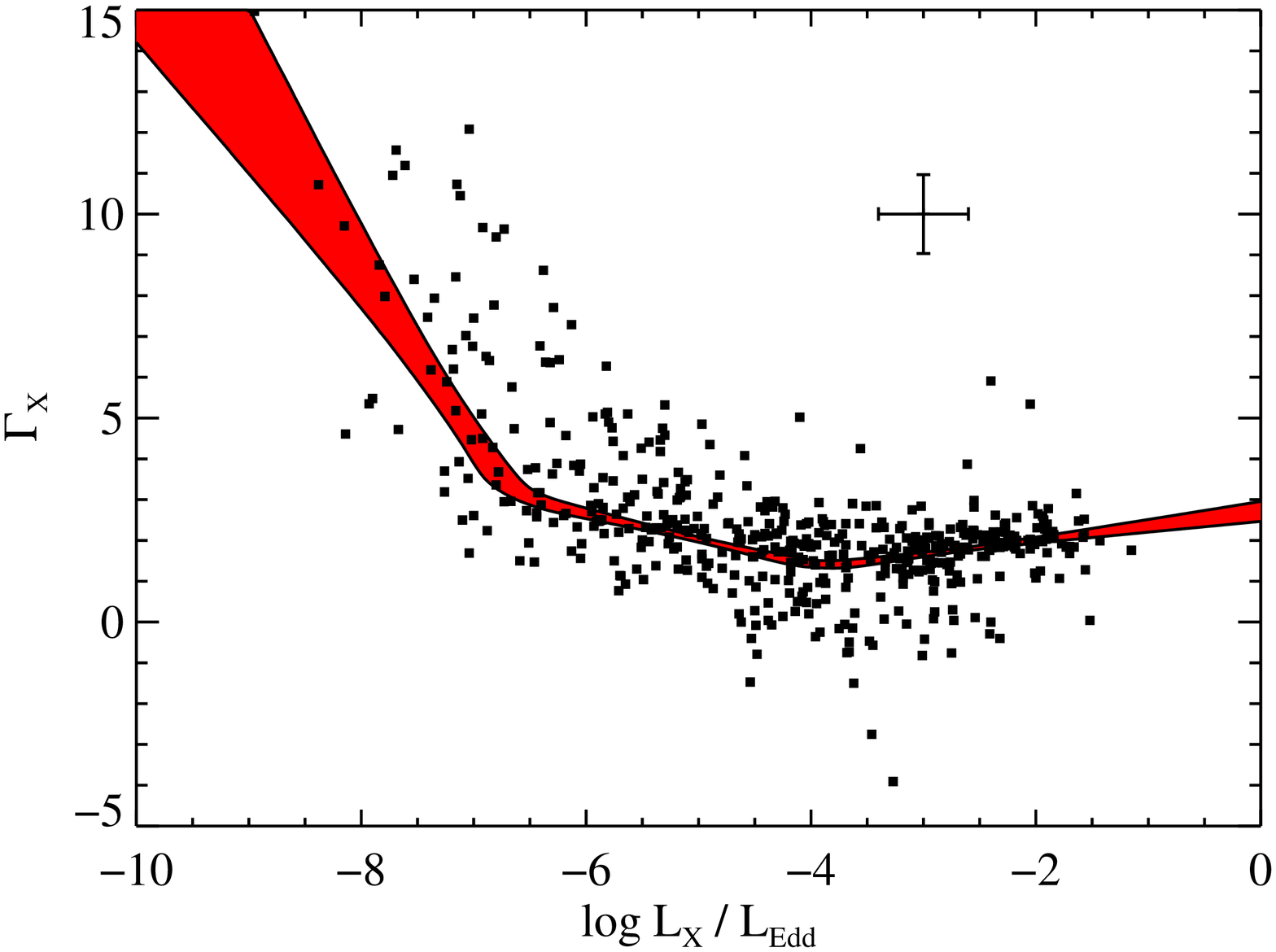}
\includegraphics[scale=0.25, angle=90]{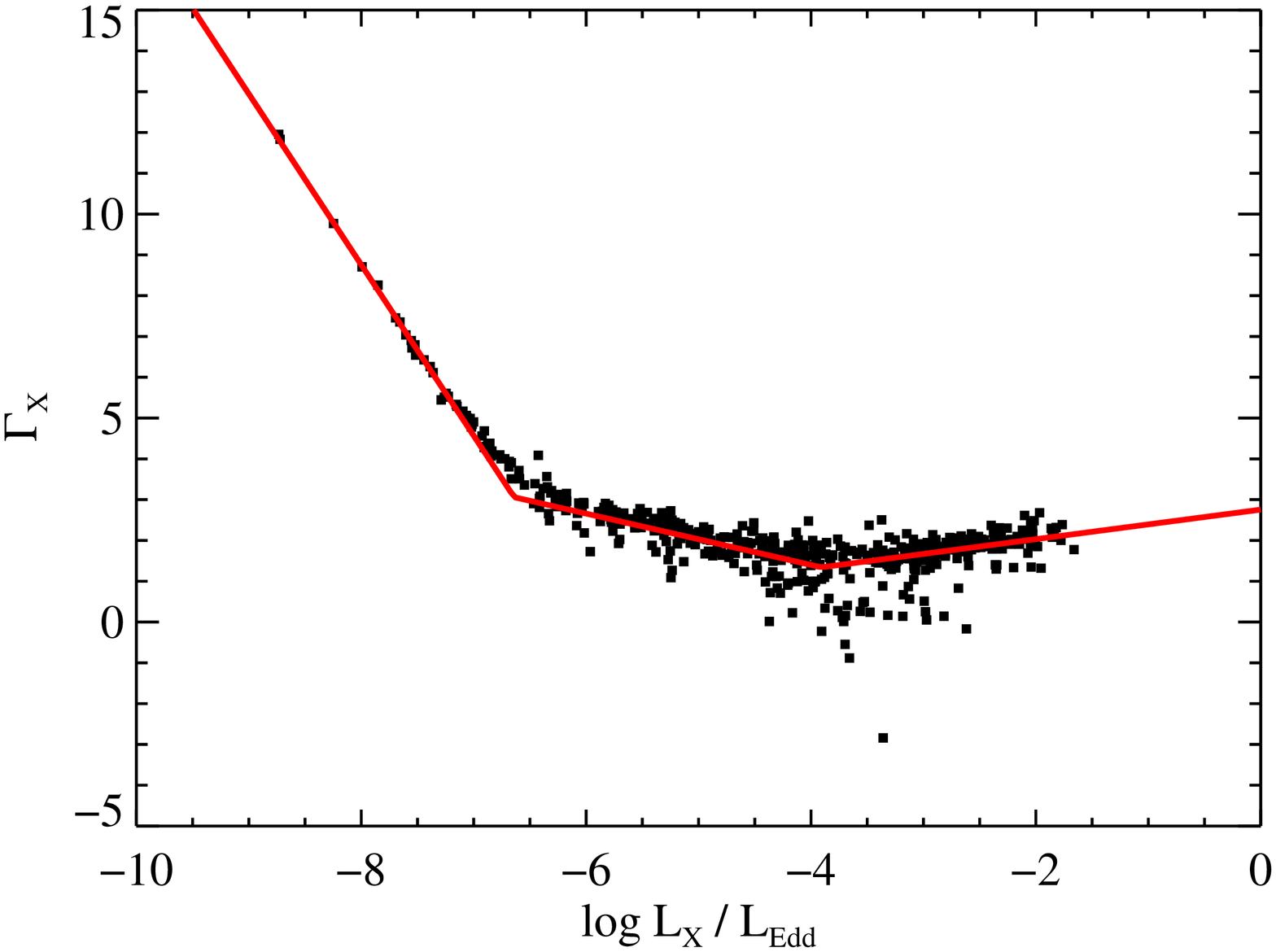}
%
%
\caption{The measured values of $\Gamma_X$ and $\log L_X / L_{Edd}$ compared with the region containing $68\%$ of the posterior probability for the mean value of $\Gamma_X$ at fixed $L_X / L_{Edd}$ (left). The data point with error bars is not real and only used to illustrate the typical size of the error bars. Also shown are the posterior mean values for the true values of $\Gamma_X$ and $\log L_{X} / L_{Edd}$, compared with the best-fitting segmented line (right). A non-linear trend is apparent in both the segmented line model and in the estimated distribution of $\Gamma_X$ and $\log L_{X} / L_{Edd}$ using the segmented line as a prior.}
\label{f-segline}       
\end{figure}

The region containing $68\%$ of the posterior probability on the mean value of $\Gamma_X$ as a function of $L_X / L_{Edd}$ is also shown in Figure 2a. The location of the knots are estimated to be $\log L_X / L_{Edd} = -6.65 \pm 0.25$ and $-3.91\pm 0.21$, respectively. The segmented line model of $\Gamma_X$ at fixed $L_X / L_{Edd}$ is preferred over a simple line model, illustrating the complex dependence of $\Gamma_X$ on $L_X / L_{Edd}$. In Figure 2b I show the posterior mean values of $\Gamma_X$ and $\log L_X / L_{Edd}$, as well as the segmented line computed from the posterior mean for its parameters. The posterior mean estimates for the true (i.e., not measured) values of $\Gamma_X$ and $\log L_X / L_{Edd}$ represent a more model-independent estimate of the dependence of the photon index on $L_X / L_{Edd}$. This represents a real advantage of the Bayesian approach, as not only are we able to estimate the probability distribution of the parameters of interest, but we can also estimate the probability distribution of the true values of the data as well, conditional on our assumed statistical model, the measured values of the data, and the amplitude of the measurement errors. The non-linear trend is also apparent from the values of $\Gamma_X$ and $L_X / L_{Edd}$ estimated from the Bayesian method. The knot at $L_X / L_{Edd} \sim 2 \times 10^{-7}$ may represent the increasing prevalence of additional astrophysical components to the X-ray spectrum as the AGN becomes faintier, such as hot gas not associated with the AGN, while the knot at $L_X / L_{Edd} \sim 10^{-4}$ may represent a change in the accretion flow geometry. Figure 2b suggest that the scatter in $\Gamma_X$ at fixed $L_X / L_{Edd}$ increases near the knot at $L_X / L_{Edd} \sim 10^{-4}$, which may be indicative of instabilities when the accretion flow changes geometry, or of uncorrected intrinsic absorption. Further analysis of this data set will be discussed in Constantin et al. (2011, in prep).

\section{Outstanding Issues in Measurement Error Models for
  Astronomical Data: Directions for Future Research} 
\label{s-future}
I will conclude by listing a couple of unsolved problems in dealing with
measurement errors in astronomical data analysis, which I hope will
lead to further research in this area.
\begin{itemize}
  \item
    {\bf Data Subject to Large, Non-Gaussian Measurement
      Errors}. Non-gaussian errors are common in astronomical data,
    especially when one is analyzing a set of derived
    quantities. Often, the most physically-interesting quantities are
    those derived by fitting an astrophysical model to the measured
    flux values at various wavelengths. Often the unertainties in
    these derived quantities are large, skewed, or exhibit multiple
    modes. There is currently no well-established method for handling
    the measurement errors in this case, although Bayesian hierarchical models
    such as that proposed by van Dyk et al. (2009) hold promise.
  \item
    {\bf Handling Measurement Errors in Massive Astronomical Data
      Sets.} Current and planned astronomical surveys will provide an
    explosion of data, allowing one to construct data sets with
    millions to billions of objects, each with multiple quantities
    measured. Many powerful methods developed for data mining will be
    applied to these data, potentially providing a powerful route to
    knowledge discovery. Unfortunately, all of the quantities obtained
    from these data sets will
    be measured with error, and most methods developed for data mining
    of massive data sets do not incorporate measurement error. This is
    especially a problem when dealing with derived quantities, which
    will likely require a more careful statistical analysis on account
    of their sometimes highly irregular error distributions. Currently,
    algorithms, such as MCMC, that allow one to perform reliable statistical
    inference on complicated statistical models do not scale well to
    massive data sets. If we want to perform inference on massive data
    sets subject to measurement error using more
    complicated and realistic statistical models, we will need advances
    on the computational side.
  \end{itemize}

  I would like to thank Anca Constantin for sharing her data set with me before publication, and Aneta Siemiginowska, Xiaohui Fan, and Tommaso Treu for helpful comments on an earlier version of this manuscript. I acknowledges support by NASA through Hubble Fellowship
  grant \#HF-51243.01 awarded by the Space Telescope
  Science Institute, which is operated by the Association of
  Universities for Research in Astronomy, Inc., for NASA, under contract
  NAS 5-26555.

%
%
%

\end{document}